\documentclass{elsart5}

 \usepackage{graphicx}

\begin{document}

\begin{frontmatter}

\title{Spin Supercurrent}
\author{Yuriy M. Bunkov\corauthref{cor1}}
\ead{yuriy.bunkov@grenoble.cnrs.fr}
\address{Centre de Recherches sur les Tr\`es Basses Temp\'eratures,
  CNRS,   38042, Grenoble, France}
\received{12 June 2005}



\begin{abstract}
 We review the main properties of Spin Waves condensation to a coherent quantum
 state, named Homogeneously Precessing Domain (HPD). We describe the long range
 coherent transport of
magnetization by Spin Supercurrent in antiferromagnetic superfluid
$^3$He. This quantum phenomenon was discovered 20 years ago. Since
then, many magnetic extensions of superconductivity and
superfluidity have been observed: spin Josephson phenomena, spin
current vortices, spin phase slippage, long distance magnetization
transport by spin supercurrents, etc. Several new supercurrent
phenomena have been discovered, like magnetically excited coherent
quantum states, NMR in the molecular Landau field, spin-current
turbulence, formation of stable non-topological solitons etc.

\end{abstract}

\begin{keyword}
\PACS 67.57.-z\sep 67.80.Jd\sep 76.50.+g

\KEY  Spin current\sep Nuclear magnetic resonance
 \sep Quantum spin liquid

\end{keyword}

\end{frontmatter}

\section{Introduction}\label{}

Quantum spin dynamics and magnetization transport are presently of a
wide interest both theoretically and experimentally. Here we
consider the magnetically ordered system, where fundamental
properties of spin quantum transport and spin waves condensation
have been discovered, superfluid $^3$He. The superfluid $^3$He is a
very interesting example of quantum liquid antiferromegnet.  In
contrast to ordinary magnetically  ordered materials where we have
spatial ordering that can be described by the language of
sublattices, here we have a mixture of quantum liquids in different
quantum states.
\begin{equation}
\Psi(\bf k) = \Psi_{\uparrow \uparrow} (\bf k)\vert \uparrow
   \uparrow \rangle + \Psi_{\downarrow \downarrow} (\bf k)\vert
   \downarrow \downarrow \rangle  + \sqrt{2}\Psi_{\uparrow \downarrow}
   (\bf k)\vert \uparrow \downarrow + \downarrow \uparrow \rangle
 \label{2.1}
\end{equation}

where  $\Psi_{\uparrow \uparrow},  \Psi_{\downarrow \downarrow}$ and
$\Psi_{\uparrow \downarrow }$ are the amplitudes
 associated with the spin substates $ \vert \uparrow \uparrow \rangle,
\vert \downarrow \downarrow \rangle $ and $\vert \uparrow \downarrow
+ \downarrow \uparrow \rangle $ respectively. The relation between
these substates can be described by the vector $\bf d$ which is
actually the axis of quantization of the Cooper pair state.

\begin{eqnarray}
\Psi (\bf k) & = &\pmatrix{\Psi_{\uparrow \uparrow} (\bf k) &
\Psi_{\uparrow \downarrow}  (\bf k) \cr \Psi_{\uparrow \downarrow}
(\bf k) & \Psi_{\downarrow \downarrow} (\bf k) \cr } \nonumber \\
& =  &\pmatrix{-d_x
 ({\bf k}) + i d_y (\bf k) & d_z (\bf k) \cr d_z (\bf k)& d_x ({\bf k})
 + i d_y (\bf k) \cr }.   \label{2.2} \end{eqnarray}

 The projection of the spin of the
Cooper pair on the direction of $\bf d$ is equal to zero,  similar
to the antiferromagnetic vector $\bf l$ in antiferromagnets.

In the case of a spatial gradient of  $ d_\perp = d_x +id_y$ the
gradients of $\Psi_{\uparrow \uparrow}$ and $\Psi_{\downarrow
\downarrow}$ have different signs. Its leads to counterflow of these
two superfluids.  This counterflow transports magnetization without
mass transport and is called the spin supercurrent.

Spin supercurrents in superfluid $^3$He have been expected for a
long time. The complexity of these phenomena follows from the fact
that it is the coherent transport of a vector quantity, not a scalar
one, as it is for mass and current charges in superfluidity and
superconductivity. Consequently the general expression for the spin
supercurrent has a tensor form and reads:

\begin{equation}
J_{i\alpha}={\hbar\over 2m}\rho_{ij\alpha\beta}\Omega _{j\beta}
\label{1.4}
\end{equation}

where $\rho_{ij\alpha\beta}$ is the spin superfluid density tensor
and $\Omega_{j\beta}$ are the phase gradients of the order
parameter.

\section{Spin Waves condensation to a coherent state}

The equations of NMR in $^3$He, so called Leggett equations, are
similar to equations of antiferromagnetic resonance with an
anisotropic term due to nuclear dipole-dipole interaction. The
solution of this equations for bulk $^3$He-B shows that the
magnetization precess on the larmore frequency if it is  deflected
less then magic dipole angle of 104$^\circ$. For bigger angles of
deflection the additional positive frequency shift appears.

In the case of NMR in a gradient of magnetic field the gradient of
$\bf d$ creates the spin supercurrent which transport the
magnetization in the direction of higher magnetic field.
Consequently the deflection of magnetization in the region of
smaller magnetic field increases up to the value of 104$^\circ$. At
further deflection, the dipole dipole frequency shift compensate the
gradient of magnetic field. The spatial magnetic energy potential
became flat and spin waves condenses to a coherent state\cite{BBDM}.
 This  state calls
Homogeneously Precessing Domain (HPD), where the magnetization is
deflected from the magnetic field and precesses
 coherently even in an inhomogeneous magnetic
 field.

  The HPD is a new type of coherent states, the
 non-equilibrium, magnetically excited state. It can be explained\cite{BTV} in
 full analogy to other coherent states  in terms  of
"off-diagonal long range ordering" ODLRO: the role of the particle
number operator is played by a quasi-conserved quantity: the
projection of magnetization ${\bf M}_z$ on the external magnetic
field. The spin creation and annihilation operators ${\bf M}_\pm$
substitute the particle-nonconserving operators. The precession
frequency $\omega$ corresponds to the chemical potential, and the
Zeeman energy $E_{Z}=-\gamma \bf H \bf M_z$ corresponds to the
particle energy $nU$ in an external scalar potential. As a result
the ODLRO in a spin system is given by
\begin{equation}<{\bf M}_-> =\sqrt{M^2-M_z^2}e^{ i(\Phi+\omega t)}
\end{equation}

It is important to note that the precession may be stable and
coherent only if the following two conditions are satisfied:

(1) The internal energy $E_D(M_z)$ (for $^3$He it is mainly the
dipole-dipole interaction) is a concave function of $M_z$ in the
same way as the concave shape of the internal energy $\epsilon (n)$
prevents the phase separation in liquid.

(2) The phase  coherence is supported by the spin rigidity: the
energy depends on the gradient of phase $E_{G}=(K/2) (\bf \nabla
\Phi)^2$, where the stiffness $K$ plays the role of the superfluid
density in mass superfluids. These conditions are satisfied for
Larmor precession in superfluid $^3$He-B in which the dipole energy
has the concave form and the magnetic stiffness is supported by the
initial stiffness of the superfluid order parameter.

The coherent HPD state is metastable, for its   creation  it is
necessarily to apply  pulsed or CW NMR techniques. In the later case
a small RF field compensates the small dissipation caused by the
non-conservation of $M_z$ due to the magnetic interaction with the
normal component of $^3$He. The RF field frequency $\omega$ plays
the role of the chemical potential and serves as the Lagrange
multiplier in the added term $E_L=- \omega M_z$. The RF field
frequency defines the equilibrium value of the tipping angle of the
precessing magnetization $M_z=M\cos\lambda$ which should correspond
to the resonance frequency $\omega$:
\begin{equation}\omega-\gamma H=-{{\partial E_D }\over {\partial M_z}}~~.\label{2.7} \end{equation}
In the same way the fixed chemical potential determines the
equilibrium particle density.

\section{Demonstration of spin supercurrent}

In order to demonstrate the non-potential flow of magnetization, we
have excited two HPD in two cylindrical cells by applying two
independent RF fields. The cells were placed in the same magnetic
field and connected by a channel of 0.6 mm diameter and 6 mm length.
The HPD states play the role of electrodes, while the RF field
frequency - the potential. In the magnetic field the spin current
transports the Zeeman energy, which can be measured by the balance
of dissipation in the two cells. We have demonstrated the spin
current flows in agreement with the magnetization precession phase
gradient along the channel. By adding an orifice inside the channel
we were able to observe the Josephson phenomenon \cite{Jos}. The
description of many quantum spin dynamics effects can be found in
the review article \cite{Bunrev}.

\section{New results}

The superfluid component of $^3$He magnetically interacts with the
normal component, which is the source of magnetic relaxation. By
cooling to very low temperatures, one can expect to observe a true
persistent NMR precession, owing to a suppression of the normal
component. Indeed, an  unpredicted instability of precession was
found at a temperatures below 0.4 T$_c$ \cite{Cat}. Now we can
explain it as a new type of Suhl instability \cite{BLV}.

At the limit of extremely low temperatures of about 100 $\mu$K, a
truly near persistent signal of NMR precession was found \cite{PIS}.
It can ring for several hours at a frequency of 1 MHz! The
theoretical and experimental investigations show, this signal is
generated by a new coherent quantum state, a non-topological soliton
\cite{PS}. This signal is sensitive to rotation, as was found in
Helsinki \cite{Hel}. Many other interesting phenomena related to
spin waves condensation and spin supercurrent can be found in
\cite{Bunrev} and current publications.


\begin{thebibliography}{00}

\bibitem{BBDM} A.~S.~Borovik-Romanov, Yu.~M.~Bunkov, V.~V.~Dmitriev,
Yu.~M.~Mukharskiy,  {\it JETP Lett.}  40 (1984), p. 1033,

\bibitem{BTV} Yu.~M.~Bunkov, O.~D.~Timofeevskaya,
G.~E.~Volovik,  {\it Phys. Rev. Lett.} 73 (1994), p. 1817.

\bibitem{Jos} A.~S.~Borovik-Romanov,  Yu.~M.~Bunkov, V.~V.~Dmitriev,
Yu.~M.~Mukhar\-skiy, D.~A.~Ser\-gatskov, {\it Phys. Rev. Lett.} 62
(1989), p. 1631.

\bibitem{Bunrev} Yu.~M.~Bunkov,  {\it Spin Supercurrent and Novel Properties of
NMR in $^3$He} in Progress in Low Temp. Physics 14, ed. W.~Halperin,
Elsevier (1995).

\bibitem{Cat} Yu.~M.~Bunkov, V.~V.~Dmitriev, Yu.~M.~Mukharsky,
J.~Nyeki and D.~A.~Sergatskov, {\it Europhys. Lett.} 8 (1989), p.
645.

\bibitem{BLV} Yu.~M.~Bunkov, V.~S.~L`vov,
G.~E.~Volovik,  {\it JETP Lett.}  83 (2006), p. 624.

\bibitem{PIS} Yu.~M.~Bunkov,  S.~N.~Fisher, A.~M.~Guenault,
G.~R.~Pickett, {\it Phys.  Rev.  Lett.} 69 (1992), p. 3092.


\bibitem{PS} Yu.~M.~Bunkov  {\it J. Low Temp. Phys} 138 (2005), p.
753.


\bibitem{Hel} Yu.~M.~Bunkov {\it et. al.}, to be published.


\end{thebibliography}
\end{document}